\documentclass[fleqn,usenatbib]{mnras}
\usepackage[T1]{fontenc}
\usepackage{ae,aecompl}

\usepackage{graphicx}	% Including figure files
\usepackage{amsmath}	% Advanced maths commands
\usepackage{amssymb}	% Extra maths symbols
\usepackage{xcolor}
\def\be{\begin{equation}}
\def\ee{\end{equation}}

\def\gsim{\lower.5ex\hbox{\gtsima}} 
\def\lsim{\lower.5ex\hbox{\ltsima}} 
\def\gtsima{$\; \buildrel > \over \sim \;$} 
\def\ltsima{$\; \buildrel < \over \sim \;$} \def\gsim{\lower.5ex\hbox{\gtsima}} 
\def\lsim{\lower.5ex\hbox{\ltsima}} 
\def\simgt{\lower.5ex\hbox{\gtsima}} 
\def\simlt{\lower.5ex\hbox{\ltsima}}

\def\Sg{$\Sigma_{\rm gas}$}
\def\S*{$\Sigma_{\rm SFR}$}
\def\Scii{$\Sigma_{\rm [CII]}$}
\def\lcii{$L_{\rm [CII]}$}
\def\Sciii{$\Sigma_{\rm CIII]}$}

\def\Soiii{$\Sigma_{\rm [OIII]}$}
\def\loiii{$L_{\rm [OIII]}$}
\def\nkZ{($\kappa_s$, $Z$, $n$)}
\def\OIIIfifty{[OIII]52$\mu$m}
\def\OIIIeighty{[OIII]88$\mu$m}

\def\HII{\hbox{H$\,\scriptstyle\rm II$}}

\def\ks{\kappa_{\rm s}}

\definecolor{apcolor}{HTML}{b3003b}
\definecolor{afcolor}{HTML}{800080}
\definecolor{lvcolor}{HTML}{0066cc}
\definecolor{mdcolor}{HTML}{01abdf} %mahsa and davide
\definecolor{cbcolor}{HTML}{ff0000}
\definecolor{sccolor}{HTML}{cc5500} %stefano
\definecolor{sgcolor}{HTML}{00cc7a}
\usepackage{deluxetable}
\defcitealias{carniani2018}{C18}
\defcitealias{ferrara2019}{F19}
\defcitealias{vallini2020}{V20}
\definecolor{apcolor}{HTML}{b3003b}
\definecolor{afcolor}{HTML}{01bdff}

\title[High \Soiii/\Scii~in the EoR]{High [OIII]/[CII] surface brightness ratios trace early starburst galaxies}

\author[Vallini et al.]{
L. Vallini$^{1}$\thanks{E-mail: livia.vallini@sns.it (LV)},
A. Ferrara$^{1}$,
A. Pallottini$^{1}$,
S. Carniani$^{1}$,
S. Gallerani$^{1}$
\\
$^{1}$Scuola Normale Superiore, Piazza dei Cavalieri 7, 56126 Pisa, Italy\\
}

\date{Accepted XXX. Received YYY; in original form ZZZ}

\pubyear{2021}

\begin{document}
%\label{firstpage}
\maketitle

\begin{abstract}
 We study the impact of deviations from the Kennicutt-Schmidt relation (quantified by the `burstiness' parameter $\kappa_s$), gas metallicity ($Z$), and density ($n$) on the observed \OIIIeighty/[CII]158$\mu$m surface brightness ratios (\Soiii/\Scii) in nine galaxies at $z\approx6-9$.
 We first discuss possible biases in the measured \Soiii/\Scii~ratios by comparing the data with zoom-in cosmological simulations, and then use a Markov Chain Monte Carlo algorithm to derive the best fit values of ($\kappa_s, Z, n$). 
 We find that (i) the strongest dependence of \Soiii/\Scii~is on $\kappa_s$; (ii) high ratios identify starburst galaxies with short gas depletion times ($t_{\rm dep}=6-49\,\rm Myr$); (iii) a secondary dependence on density is found, with \Soiii/\Scii~ anticorrelating with $n$ as a result of the lower [OIII] critical density, (iv) the ratio only weakly depends on $Z$. The nine galaxies are significantly enriched ($Z=0.2-0.5 Z_\odot$), and dense $n \approx 10^{1-3} {\rm cm}^{-3}$. This lends further support to the  starburst scenario in which a rapid enrichment of the interstellar medium is expected.
  
\end{abstract}
%%%%%%%%%%%%%%%%%%%%%%%%%%%%%%%%%%%%%%%%%%%%%%%%%%
%%%%%%%%%%%%%%%%% BODY OF PAPER %%%%%%%%%%%%%%%%%%
\section{Introduction}

The Atacama Large Millimeter Array \citep[ALMA;][]{carilli2013} opened a window on the characterization of the interstellar medium (ISM), star formation, and chemical enrichment in the Epoch of Reionization (EoR) as traced by far-infrared (FIR) lines.
Among FIR lines, the fine-structure ${}^2P_{3/2}\rightarrow{}^2P_{1/2}$ transition of the ionized carbon ([CII]) at 158$\mu$m is one of the most luminous \citep{stacey1991}, with the [CII] line mostly tracing the cold neutral diffuse gas \citep{wolfire2003} and the dense photodissociation regions \citep[PDRs;][]{hollenbach1999} associated with molecular clouds.
After the first, pioneering, detections in galaxies at $z>5.5$ \citep{capak2015, maiolino2015}, [CII] has now been observed and often spatially resolved, in $\approx$100 galaxies at $4<z<5.5$ \citep[ALPINE-Survey;][]{lefevre2020, bethermin2020}, and in tens of $z=6-8$ sources \citep[][]{willott2015, knudsen2016, pentericci2016, bradac2017, matthee2017, smit2018, carniani2018, carniani2018b, hashimoto2019, fujimoto2019, matthee2019, bakx2020, calura2021, herrera-camus2021}.
In addition to [CII], ALMA has been also exploited to target EoR sources via the ${}^3P_{1}\rightarrow {}^3P_{0}$ transition of doubly ioni\-zed oxygen ([OIII]) at 88$\mu$m \citep{inoue2016, carniani2017, laporte2017b, hashimoto2019, tamura2019, harikane2020, carniani2020}. \\

Joint [CII]-[OIII] line detections have a huge diagnostic potential as they yield complementary views of the ISM at early epochs. While [OIII] traces ionized gas in \HII~ regions \citep[][]{cormier2015}, [CII] mainly arises from neutral/molecular gas. 
Notably, the [OIII]/[CII] luminosity ratios (\loiii/\lcii) in galaxies at $7<z<9$ \citep{inoue2016, hashimoto2019, harikane2020, carniani2020} exceed the highest values observed in local dwarf galaxies \citep{madden2013, cormier2015} which are known to be bright [OIII] emitters.

An increasing number of theoretical works \citep{vallini2013, vallini2015, olsen2017, pallottini2017b, katz2017,lagache2018, kohandel2019, pallottini:2019, lupi2020b, arata2020}, suggest that the prevailing physical conditions of the ISM were extreme (e.g. large densities, high turbulence, strong radiation fields) and common among early galaxies. These findings provide a solid basis to investigate the origin of the observed high \loiii/\lcii~ratios.

Possible explanations include high ionization parameters ($U$) \citep{katz2017, moriwaki2018,pallottini:2019}, high filling factor of ionized gas vs gas in dense PDRs \citep{harikane2020}, intermittent starbursting phases in the star formation histories of the EoR galaxies \citep{arata2020, lupi2020b}, and photoevaporation feedback \citep{vallini2017, decataldo2017}. As outlined by \citet{ferrara2019} all these conditions can be produced in the ISM of a starburst galaxy, hence located above the Kennicutt-Schmidt (KS) relation. 
As a matter of fact, a source with star formation rate surface density (\S*) exceeding the the KS value at fixed gas surface density (\Sg), has a correspondingly larger $U$, and therefore larger ionized gas column density as compared to a galaxy with the same \Sg~but lying on the KS relation. These conditions boost (quench) ionized (neutral) gas tracers \citep{vallini2020}.

Apart from gas ionization conditions, [CII] and [OIII] line emission is also influenced by metallicity ($Z$):  it has been shown that for $Z < 0.1\, \rm Z_{\odot}$ the [CII] luminosity drops significantly \citep[e.g.][]{vallini2015, pallottini2017b, olsen2017, lagache2018}. At the same time, low-$Z$ galaxies are expected to feature brighter [OIII] luminosities \citep{cormier2015, inoue2016, vallini2017}. Lower C/O ratios \citep{steidel2016, arata2020}, high densities \citep{harikane2020}, and the compactness of EoR galaxies \citep{shibuya2015} can also partially explain large \loiii/\lcii~ ratios.

However, the determination of \loiii/\lcii~ratios in early galaxies must be handled with caution. As noted by \citet{carniani2020} the \loiii/\lcii~values of EoR sources are closer to those of local metal-poor dwarf galaxies when the [CII] flux loss is corrected for. The \loiii/\lcii~are in fact prone to overestimation because the [CII] emitting region of ALMA detected galaxies at high-$z$ is $\approx2-3$ times more extended \citep{carniani2018b,fujimoto2019, ginolfi2020, fujimoto2020, herrera-camus2021} than the [OIII]/rest-frame ultraviolet (UV) one. Flux losses due to the surface brightness dimming in [CII] can thus lead to a spurious underestimation of the actual [CII] luminosity \citep{kohandel2019}.

Analyzing instead the [OIII]/[CII] surface brightness ratios (\Soiii/\Scii) can overcome this problem as the different extension of the line emitting regions are explicitly accounted for when computing the line surface brightness. Additionally, the \Soiii/\Scii~ratios, along with their relation with \S*, are more closely related to the \emph{local} ISM conditions than the integrated \lcii/\loiii~ratios. For instance, the far-ultraviolet flux impinging upon PDRs -- which is one of the fundamental the parameters affecting the line surface brightness  from PDR and ionized gas layers \citep[e.g.][]{ferrara2019, vallini2020} -- is more tightly related to the local SFR surface density   \citep[e.g.][]{herrera-camus2015, diaz-santos2017, rybak2020, mckinney2021} than to the total galaxy SFR.\\

The aim of this work is precisely that of exploiting \Soiii/\Scii~ratios with the goal of determining the star-forming and chemical enrichment conditions of the ISM in EoR galaxies, along with the physical mechanisms governing the [OIII] and [CII] emission in early sources. To this aim, we build on the \citet{ferrara2019, vallini2020} models which provide a physically motivated framework to compute the expected FIR line surface brightness tracing both ionized and neutral gas as a function of the deviation from the KS relation ($\kappa_s$), metallicity (Z), and gas density ($n$). The model also provides physically transparent interpretations of complex numerical simulations \citep[][]{pallottini:2019}. The paper is organized as follows: in Sec. \ref{sec:method} we present the method and validate it on well-studied low-$z$ sources; in Sec. \ref{sec:application_and_comparison} we apply the model to all the joint [CII]-[OIII] detections in EoR so far available and compare the results with state-of-art cosmological zoom-in simulations. Sec. \ref{sec:results} discusses the implications of the results; our conclusions are given in Sec. \ref{sec:conclusions}.

\section{Method}\label{sec:method}
To study the physical mechanisms determining the \Soiii/\Scii~ratios
we adopt the method presented in \citet[][V20 hereafter]{vallini2020}. In  \citetalias{vallini2020} the [CII] surface brightness (\Scii), the SFR surface density (\S*), and the surface brightness of a ionized gas tracer ($\Sigma_{\rm line,ion}$), have been used to determine the (possible) deviation from the star formation law, the gas density and metallicity of galaxies in the EoR. \citetalias{vallini2020} focused on $\Sigma_{\rm line,ion}=$\Sciii, i.e. the UV CIII]$\lambda1909$ line, which is expected to be bright in the first galaxies \citep[e.g.][]{stark2015}. For the present analysis we apply the model to the \OIIIeighty~surface brightness (\Soiii). In what follows we summarize the rationale of the procedure, the fundamental equations, and  their extension to oxygen lines.
 
\subsection{Rationale, fundamental equations, and extension to [OIII]}
The \citetalias{vallini2020} method is built upon the analytical model de\-ve\-lo\-ped in \citet{ferrara2019} that enables the computation of the line surface brightness from a gas slab with ionized/PDR column densities ($N_{\rm PDR}$, $N_i$, respectively) determined by the average gas density ($n$) of the $\HII$/PDR environment, the dust-to-gas ratio, (${\cal D} \propto Z$), and ionisation parameter, $U$. The latter, can be expressed in terms of observed quantities by deriving its relation \citep[$U\propto \Sigma_{\rm SFR}/\Sigma_{\rm gas}^2$, see eq.s 38 and 40 in][]{ferrara2019} with the star formation rate surface density (\S*) and the gas surface density (\Sg), which in turn are connected through the star formation law $\Sigma_{\rm SFR} = \kappa_s \Sigma_{\rm gas}^{1.4}$ \citep{kennicutt1998}.
This leaves us with the $\kappa_s$ parameter, describing the burstiness of the galaxy.

\citetalias{vallini2020} adopted a Markov Chain Monte Carlo (MCMC) algorithm to search for the posterior probability of the best fit parameters \nkZ~that reproduce the observed [CII] surface brightness ($\Sigma^{obs}_{\rm [CII]}$, in $L_{\odot}\,\rm kpc^{-2}$ units),  CIII]$\lambda$1909 surface brightness ($\Sigma^{obs}_{\rm CIII]}$, in $L_{\odot}\,\rm kpc^{-2}$), and the deviation ($\Delta^{obs}_{\rm [CII]}$) from the local \Scii-\S*~relation \citep{delooze2014}:
\begin{eqnarray}
     \label{eq:scii}
    &\Sigma^{obs}_{\rm [CII]} = F_{\rm [CII]} (\ks, Z, n)\\
    \label{eq:sion}
    &\Sigma^{obs}_{\rm CIII]} = \, F_{\rm CIII]} (\ks, Z, n)\\
    \label{eq:delta}
    &\Delta^{obs}_{\rm [CII]} (\Sigma^{obs}_{\rm SFR}) = \Delta_{\rm [CII]}(\ks, Z, n)
\end{eqnarray}
\begin{figure*}
    \centering
    \includegraphics[scale=0.41]{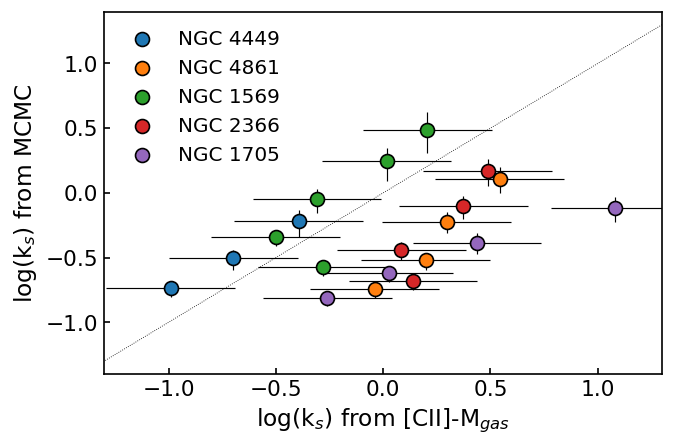}
    \includegraphics[scale=0.41]{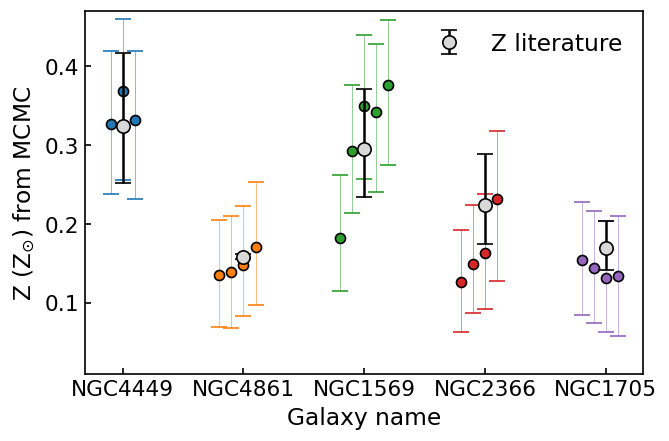}
    \includegraphics[scale=0.41]{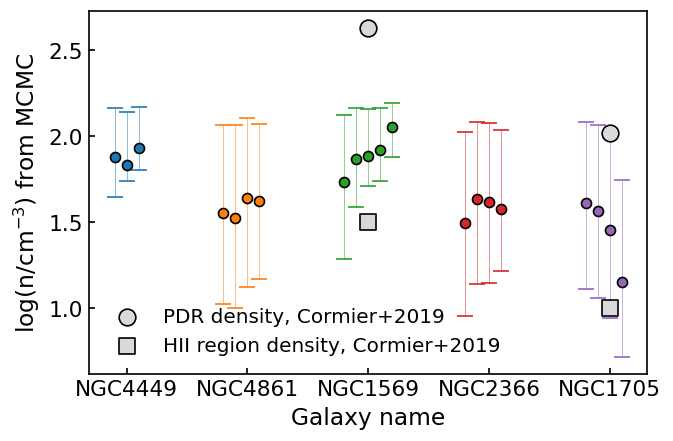}
    \caption{\emph{Left:} Burstiness parameters derived with our model vs those derived from [CII] surface brightness by exploiting the \citet{madden2020} relation. Each galaxy (color-coded) has multiple $\kappa_s$ estimates corresponding to different \S*~bins. \emph{Center:} Scatter plot showing, for each galaxy, the best-fit value of $Z$ and $1\sigma$ errors returned by the MCMC in different \S*~bins. Grey points with errorbars represent the metallicity derived from the 12+log(O/H) values reported in the literature: NGC 4449 and NGC 4861 \citep{madden2013}, NGC 1569 \citep{mccormick2018}, NGC 2366 \citep{james2016}, and NGC 1705 \citep{annibali2015}. \emph{Right:} Scatter plot showing, for each galaxy, the gas density from MCMC and the $1\sigma$ errors for the various \S*~bins. For comparison, grey dots (grey squares) show the PDR (\HII~region) density derived by \citet{cormier2019} in NGC 1569 and NGC 1705.}
    \label{fig:comparison}
\end{figure*}
In this work, instead of Eq. \ref{eq:sion} we use $F_{\rm [OIII], 88}(\ks, Z, n)$, so that the input of the MCMC is the \OIIIeighty~surface brightness ($\Sigma^{obs}_{\rm [OIII]}$) instead of $\Sigma^{obs}_{\rm CIII]}$. To do so, we follow the same approach outlined in Eq. 29 and in Appendix B of \citet{ferrara2019}. Adopting the same notation used in \citetalias{vallini2020}, 
the $F_{\rm [OIII],88}$ flux (in erg s$^{-1}\, \rm cm^{-2}$) excited by collision with free electrons in a gas slab of density $n$ and ionized hydrogen column density $N_i(\ks, Z)$, can be written as a function of the cooling rate ($\Lambda_{\rm [OIII], 88}$) as:
\begin{equation}
F_{\rm [OIII],88} = n \Lambda_{\rm [OIII],88} {\cal A}_{O} {Z}  N_i(\ks, Z).
\label{eq:foiii88}
\end{equation}
In the previous equation the O$^{2+}$ column density is approximated as $N_{O^{2+}} \approx {\cal A}_O Z N_i$ where we assume \citet{asplund2009} abundance at solar metallicity (${\cal A}_O\equiv {\rm O/H} = 4.89 \times 10^{-4}$ or $12+\log ({\rm O/H})=8.69$\footnote{The carbon abundance entering the [CII] emission calculation is ${\cal A}_C\equiv {\rm C/H} =2.69\times 10^{-4}$ or $12+\log ({\rm C/H})=8.43$}). Moreover, we adopt a linear scaling with $Z$ (which is in solar units). 

The cooling rate ($\Lambda_{\rm [OIII], 88} = n_1 A_{10} E_{10}$) follows from the computation of the population of the $^3P_1$ level. To derive the cooling rate as a function of temperature, $T$, and density $n$, we use \texttt{Pyneb} \citep{luridiana2015}. Such code  solves the statistical equilibrium equation for the O$^{2+}$ ion including all the possible transitions between the $^3P_0,^3P_1,^3P_2,^1D_2,^1S_0$ levels: 
 \begin{equation}
  \sum_{j\neq i} n n_j C_{ij}(T) +  \sum_{j> i} n_j A_{ij}= \sum_{j\neq i} n_j n C_{ij} (T) +  \sum_{j<i} n_i A_{ij}.
\end{equation}
where $C_{ij}$ are the collisional excitation (de-excitation) rate coefficients, $A_{ij}$ are the Einstein coefficients for spontaneous emission, and $i=0,.., 4$. The electron temperature in the ionized region is set to $T= 10^4\, \rm K$. This assumption is not very critical as the $^3P_1 \rightarrow ^3P_0$ \OIIIeighty~ line (and the other transition in the doublet, i.e. the $^3P_2 \rightarrow ^3P_1$ transition at 52$\mu m$) have similar excitation energy ($T_{{\rm ex}, 88}\approx 160\,\rm K$ and $T_{{\rm ex}, 52}\approx 280\, \rm  K$, respectively) but they have dif\-fe\-rent critical densities $n_{c}\equiv A_{ij}/C_{ij}$. For this reason the [OIII]88$\mu$m/[OIII]52$\mu$m ratio is not very sensitive to the gas temperature for $T>1000 \, K$ \citep{palay2012}. \\

Eq.s \ref{eq:scii}, \ref{eq:foiii88}, \ref{eq:delta}, and the observed values and errors, $\Sigma^{obs}_{\rm [CII]} \pm \delta_{\rm [CII]}$, $\Sigma^{obs}_{\rm [OIII]}\pm \delta_{\rm [OIII]}$, $\Sigma^{obs}_{\rm SFR}\pm \delta_{\rm SFR}$ for each galaxy, allow us to search for the posterior probability of the best fit parameters exploiting a MCMC algorithm. As in \citetalias{vallini2020}, we use the open-source \texttt{emcee} Python implementation \citep{foreman2013} of the Goodman Weare’s Affine Invariant MCMC Ensemble sampler \citep{goodman2010} adopting flat priors in the range $0.0 \leq \log n \leq 4.0$, $-2 \leq \log Z \leq 0.0$ and $-1 \leq \log k_s \leq 2.5$, i.e. spanning all the physically reasonable parameter space\footnote{We publicly release the code, called GLAM (Galaxy Line Analy\-zer with MCMC), on GitHub  (\url{https://lvallini.github.io/MCMC_galaxyline_analyzer/}) along with Jupyter notebooks that exemplify how to derive the \nkZ~of any galaxy of interest for which \Scii, \Soiii, and \S* are measured.}. We use the $\chi^2$ likelihood function to determine the probability distribution function of the output parameters.

\begin{figure*}
    \centering
    \includegraphics[scale=0.57]{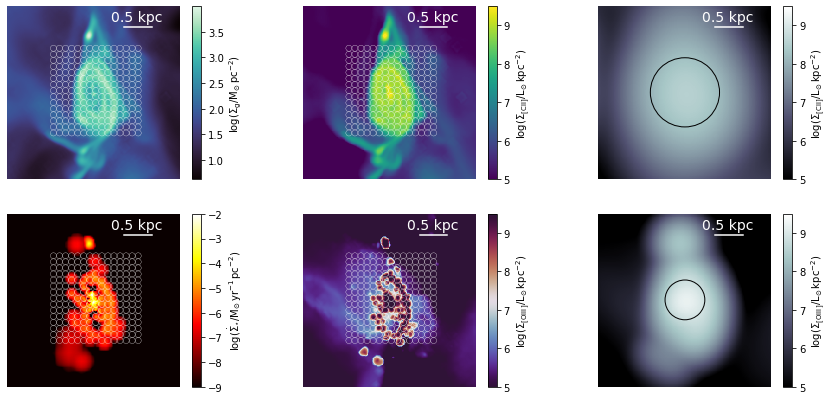}
    \caption{Cutouts of \textit{Iris} (SERRA-04:46:4630), one the 20 galaxies extracted from the SERRA simulation \citep{pallottini2021}. We also overplot the circular patches of $d=\rm 100\, pc$ covering a square of 1.5 kpc edge, used for our spatially resolved analysis (see text for details).
    \textit{Upper row:} gas surface density (left), [CII] surface brightness (center) at 30 pc resolution, and that obtained after convolving with $\theta=0.2''$ beam (right). The half light radius is highlighted with a black circle. 
    \textit{Bottom row:} SFR surface density map (left), [OIII] surface brightness at 30 pc resolution, and that obtained after convolving the simulation with a $\theta = 0.1''$ beam (right). The half light radius is indicated with a black circle.}
    \label{fig:SERRA_example}
\end{figure*}

\subsection{Model validation at low redshift}
\label{sec:validation}
Before applying our model to galaxies in the EoR we validate it on a sub-sample of five spatially resolved sources extracted from the Dwarf Galaxy Survey \citep[DGS,][]{madden2013}. Dwarf galaxies with their bursty star formation activity, low-metallicity, and low stellar mass, are generally considered fair local analogues of high-$z$ sources \citep[e.g.][]{cormier2012, ucci2019, nanni2020}. The five galaxies analy\-zed here are NGC 4449, NGC 4861, NGC 1569, NGC 2366, NGC 1705 for which we consider the spatially resolved HERSCHEL \Scii, \Soiii, and GALEX FUV, and MIPS 24 $\mu$m measurements (converted to \S*) computed by \citet{delooze2014} over pixels of physical size $114 \times 114$ pc$^2$.\footnote{Only pixels attaining surface brightness levels of signal-to-noise $S/N>5$ were taken into consideration by \citet{delooze2014}.} For each galaxy, we aggregated the pixels in 0.5 dex bins in the range $\log (\Sigma_{\rm SFR}/{\rm M_{\odot}\, yr^{-1}\, kpc^2}) =[-3, 0.5]$, and associated to each of those bins the mean value of \Scii, \Soiii\footnote{The result of this procedure is detailed in Appendix \ref{appendixA}}. These values are then fed as input to the model to compute the likelihood distribution of the \nkZ~parameters over each bin.

In Figure \ref{fig:comparison}, for each galaxy we compare the \nkZ~ va\-lu\-es with independent estimates in literature. The left panel compares $\kappa_s$ values obtained from our model with those estimated as $\kappa_s \propto \Sigma_{\rm gas}^{1.4}/\Sigma_{\rm SFR}$, i.e. by inverting the KS relation. This is done by inferring the \Sg~in each \S*~bin from the \Scii~using $L_{\rm [CII]}-M_{gas}$ conversion factor in the DGS  \citep{madden2020}. Note that this assumption is a rather uncertain as the $L_{\rm [CII]}-M_{gas}$ relation involves integrated, rather than areal, quantities. We find that our $\ks$ estimates for NGC 4449 and NGC 1569 are in agreement within errors with those inferred by using the \citet{madden2020} conversion factor. However, for NGC1705, NGC 4861, and NGC 2366, $50\%$ of the bins show lower $\kappa_s$ values with respect to those inferred from the $L_{\rm [CII]}-M_{gas}$ conversion, falling at $\approx 1.5 \sigma$ from the value estimated using the \citet{madden2020} relation. This is likely because the errors on the $\kappa_s$ derived applying the \citet{madden2020} conversion consider only the uncertainty on the $L_{\rm [CII]}-M_{gas}$ relation, thus they are likely underestimated. In fact, we do not include other systematics inherent to the conversion from integrated to surface brightness values (e.g. contamination from the ionized gas phase partially contributing to the [CII] emission along the line of sight).
In the central panel of Fig \ref{fig:comparison}, for each galaxy, we compare the metallicity obtained in each (\S*, \Scii, \Soiii) bin, with global values reported in literature \citep{madden2013, mccormick2018, james2016, annibali2015}. We find excellent agreement as the different $Z$ inferred in the various bins of each source are within the errors of the global galactic values.
Finally, in the right panel of Fig \ref{fig:comparison} we compare the gas density inferred from our model over each (\S*,\Scii,\Soiii) bin with literature values. This test can be performed only in two out of five sources (NGC 1569, NGC 1705) for which the average \HII~and PDR gas densities are available \citep{cormier2019}. \citet{cormier2019} derived the electron and PDR densities by simultaneously fitting mid-infrared lines from ionized gas ([ArII], [ArIII], [SIV], [NeII], [NeIII], [SIII], [NII], [NIII], [OIII]) and far-infrared neutral gas tracers ([SIII], [OI], [CII]) with CLOUDY models accounting for both phases and for the ionized-neutral boun\-da\-ry transition. Despite the caveat that systematic biases can be present when using different methods, we find a nice agreement between our values and those retrieved by \citet{cormier2019}. We note that the \HII~and PDR densities from \citet{cormier2019} enclose the lower and upper bounds, respectively, of the density distribution derived with our method. This is expected as our $n$ value represents the average density of the \HII/PDR  environment from which [OIII] and [CII] line emission arises.

\section{Applying the model in the EoR}
\begin{figure*}
    \centering
    \includegraphics[scale=0.75]{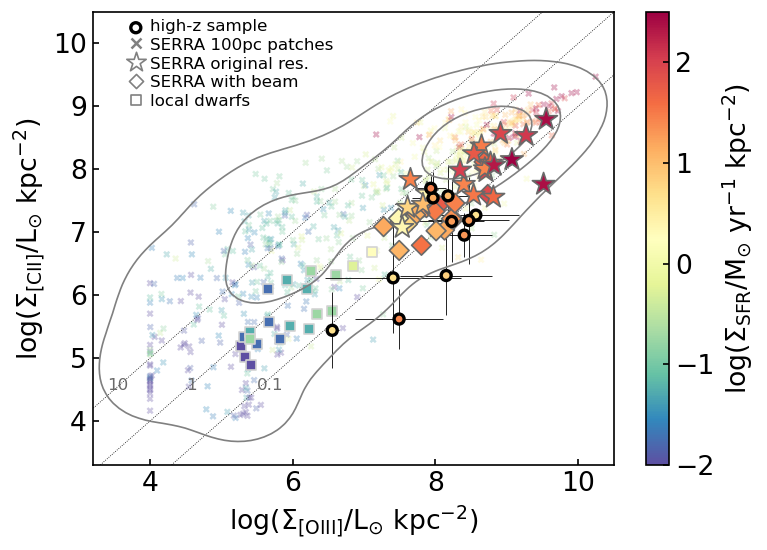}
    \caption{\Soiii~vs \Scii~relation for the high-$z$ galaxies analysed in this work (circles with error-bars) color-coded as a function of \S*. The stars represent the result for the 20 sources extracted from SERRA at $z=7$ at 30 pc resolution. The results obtained when convolving the simulation with the ALMA beam are instead plotted with diamonds. Small crosses represent values extracted from 100 pc scale ISM patches within SERRA galaxies. The $1\sigma$, $2\sigma$, $3\sigma$ levels of the patches distribution are plotted in grey. The  $z=0$ dwarf galaxy sample is denoted by squares. To guide the eye, we indicate the 1:1 relation  (and the $\pm 1 \rm dex$ dispersion around it) with dotted black lines}
    \label{fig:soiii_scii_ssfr}
\end{figure*}
\label{sec:application_and_comparison}
More than 30 EoR galaxies have been so far detected, and often spatially resolved in [CII] \citep[e.g.][for compilations]{carniani2018, matthee2019} and nine very bright galaxies (see Table \ref{tab:higzsources}) at $z\approx 6-9$ have been detected in [OIII]. The list includes MACS1149-JD1 \citep{hashimoto2018}, A2744-YD4 \citep{laporte2017}, MACS416-Y1 \citep{tamura2019}, SXDF-NB1006-2 \citep{inoue2016}, B14-65666 \citep{hashimoto2019}, BDF3299 \citep{carniani2017}, J0121 J0235, J1211 \citep{harikane2020}. All of them have been detected also in [CII], albeit at fainter fluxes with respect to [OIII] \citep[][]{carniani2020, bakx2020, hashimoto2019, maiolino2015, harikane2020}, and therefore are suitable for use in our model.

From the line luminosity, and the sizes of the emitting regions ($r_{\rm [CII]}$, $r_{\rm [OIII]}$) obtained by \citet{carniani2020}, we compute the surface brightness of each line as \Scii$=L_{\rm [CII]}/2\pi r_{\rm [CII]}^2$, and \Soiii$=L_{\rm [OIII]}/2\pi r_{\rm [OIII]}^2$. \S* is inferred either from the UV-derived star formation rate (SFR$_{\rm UV}$) or, for sources detected in continuum by ALMA (A2744-YD4, MACS0416-Y1, B14-65666, J1211 and J0217), from the UV+IR luminosity  \citep[SFR$={\rm SFR}_{\rm UV} + {\rm SFR}_{\rm IR}$,][]{carniani2020} using the \citet{kennicutt2012} relations. In particular \citet{carniani2020} derived the IR luminosity using a modified blackbody with dust temperature $T_d = 40\, \rm K$ and emissivity index $\beta = 1.5$ to reproduce the observed continuum measurements.
The area of the star forming region is taken equal to the UV emitting region, $\pi r_{\rm UV}^2$, following \cite{carniani2018}, and  \citet{vallini2020}. We stress two important caveats: (i) there is growing evidence that a large fraction of the high-$z$ galaxy population is characterized by a multi-component morphology \citep[e.g.][]{maiolino2015, matthee2017, jones2017, carniani2018, hashimoto2019}, and (ii) this often leads to significant spatial offsets (up to several kpc) between [CII], [OIII], dust continuum emission, and the star-forming regions traced by rest-frame UV light \citep[][]{carniani2017, bakx2020}.
Among the galaxies in our sample,  B14-65666 and BDF3299 show a clear multi component morphology, MACS416-Y1 and BDF3299 have spatial offsets between [CII] and [OIII]. B14-65666 has been spatially resolved in two clumps (A and B) which are both detected in [CII], [OIII] and IR continuum \citep{hashimoto2019}. This allow us to apply our model to B14-5666 and its two clumps separately. Doing the same for BDF3299 is impossible because the [CII] and the [OIII] are not co-spatial. Hence we warn the reader that for BDF3299 and MACS416-Y1 we use the global \Scii,\Soiii, and \S* not considering the spatial displacement, an information that would require an additional parameter in our model. 

The mean \Soiii/\Scii~ratio of the EoR sample is $\approx 10\times$ higher than observed locally. As we will see in the next Section, this has strong implications for the predicted \nkZ~values. Hence in the following we pause to discuss whether such high values may be affected by observational biases due to beam smearing \citep[][]{kohandel2019, kohandel2020} or if instead they are produced by extreme conditions in the ISM of early galaxies. 

\subsection{Possible biases affecting high \texorpdfstring{$\Sigma_{\rm [OIII]}/\Sigma_{\rm [CII]}$}{X} values}
To study the \Soiii/\Scii~expectation from a theoretical point of view, we consider a sample of 20 galaxies at $z=7.7$ extracted from SERRA\footnote{SERRA follows the evolution of galaxies down to $z=6$ with mass (spatial) resolution of the order of $10^{4}M_{\odot}$ (30 pc at $z\approx6$). The simulation adopts a multi-group radiative transfer version of the hydrodynamical code RAMSES \citep{teyssier:2002, rosdahl2013} that includes thermochemical evolution via KROME \citep{grassi:2014mnras, bovino:2016aa, pallottini2017b}, which is coupled to the radiation field \citep{pallottini:2019, decataldo2020}, a state-of-the-art cosmological zoom-in simulation presented in \citet{pallottini2017b, pallottini:2019, pallottini2021}  Stellar feedback includes SN explosions, OB/AGB winds, and both in thermal and turbulent form \citep{pallottini2017b} The [CII] and [OIII] emission is obtained by post-processing the outputs with CLOUDY model grids \citep[c17.0 version,][]{ferland2017} accounting for the internal structure of molecular clouds \citep{vallini2017,pallottini:2019}.}. The sample extracted from SERRA is built by selecting central galaxies with stellar mass in the range $\log (M_*/M_{\odot})=8.0-10.1$, comparable to that spanned by the observed $z=7-9$ sources in our EoR sample  \citep{roberts-borsani2020, jones2020}.

In Fig. \ref{fig:SERRA_example} we show the cutouts of \emph{Iris}\footnote{Galaxies extracted from SERRA (``greenhouse" in Italian) are named after flower species.}, the most extended source in our sample.
To perform a fair comparison with the [OIII] and [CII] observations in low-$z$ dwarf galaxies, we first derive \Scii, \Soiii, \S*~within circular patches of $100\,\rm pc$ diameter covering an area of $1.5\,\rm kpc\times1.5\, kpc$ centered on each galaxy (see white circles in Fig. \ref{fig:SERRA_example}). Considering that typical UV size of sources at $z=6-7$ is $\approx 0.5-1.0$ kpc \citep{shibuya2015, shibuya2019} and that the [C II] half-light radius is typically $r_{\rm [CII]}\approx 2.5 \pm 1.0$ \citep{carniani2018, matthee2019}, the region covered by the patches is wide enough to include the ISM outskirts (see Fig. \ref{fig:SERRA_example}). We also compute the global \Scii, \Soiii, \S*~values in two different ways: (i) considering the (native) 30 pc resolution galaxy maps from SERRA, and dividing the [OIII] ([CII]) luminosity of each galaxy at the half-light radius, by the corresponding area, (ii) convolving the [OIII] ([CII]) maps with a gaussian kernel of 0.1 arcsec (0.2 arcsec) resolution and following the same procedure. The resolutions chosen are equal to the lower limits for the beam adopted in the observations analyzed by \citet{carniani2020}. In both cases \S* is the mean value computed at the half light radius of the [OIII] emission. 

The results of this analysis are shown in Figure \ref{fig:soiii_scii_ssfr}. There, we plot  \Scii~vs \Soiii~values for the $100\, \rm pc$ patches, and the galaxy-averaged values computed both at the 30 pc SERRA resolution, and after convolution with the ALMA beam. The surface brightness of individual patches span a wide range of down to \Soiii~$\simeq$ \Scii $\approx 10^{5}-10^{6}\, \rm L_{\odot}\, kpc^{-2}$, a value similar to that observed in local dwarfs. These faint regions are characterized by low \S*$\approx 10^{-2}\rm M_{\odot}\,yr^{-1}\, kpc^{-2}$. However, these patch properties are not common in $z=7$ galaxies, they deviate by $\simgt 2\sigma$ from the (luminosity weighted) mean which is located at much higher \Scii$\approx 10^{8}\rm L_{\odot}\, kpc^{-2}$, \Soiii$\approx 10^{8.5}\rm L_{\odot}\, kpc^{-2}$ surface brigthness, or \S*$\approx 10^{1.5}\rm M_{\odot}\,yr^{-1}\, kpc^{-2}$. This confirms that early sources are highly star-forming, with correspondingly bright [CII], [OIII] arising from \HII~regions and dense PDR \citep[][]{hollenbach1999, vallini2015} associated with Giant Molecular Clouds (GMCs), in agreement with finding from other groups \citep{katz2017, arata2020}.

 Overall, 50\% (85\%) of the \Soiii, \Scii, \S*~galaxy-averaged values, at native resolution, fall within the 1$\sigma$ (2$\sigma$)  dispersion of the patches. In particular, their distribution peaks around the luminosity-weighted mean of the patches.
 The beam-convolved data shifts towards slightly lower \Scii,\Soiii~and \S*. However, the \Soiii/\Scii~ratio remains constant as both the beam-convolved and the non-convolved values are $\approx 5-10 \times$ higher than those measured in the local dwarfs on sub-kpc scales. The luminosity weighted \Soiii/\Scii~ values on global galactic scales in the dwarf galaxies could potentially shift towards higher values thus coming closer to those derived in the high-$z$ sample. Nevertheless, from our simulation, we do not obtain any systematic shift of global values towards parameter regions that are not covered by \emph{existing} patches hence, we do not expect EoR observations to be be affected by spurious bias effects.
 \begin{figure*}
    \centering
     \includegraphics[scale=0.62]{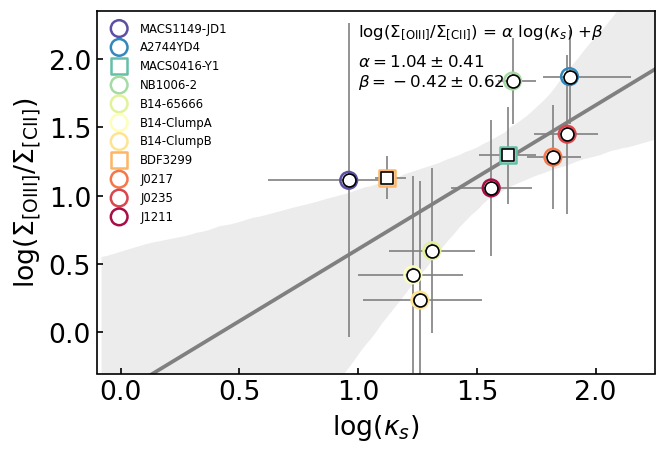}
    \includegraphics[scale=0.62]{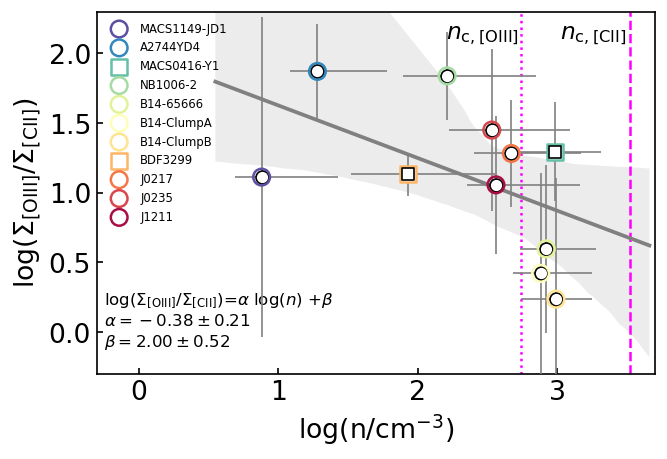}
    \caption{\emph{Left panel:} Best-fit log(\Soiii/\Scii)-$\log \kappa_s$ relation and its $1\sigma$ dispersion (black line and shaded area) for the high-z joint [CII]-[OIII] sample analysed in this work.
    \emph{Right panel:} Best fit log(\Soiii/\Scii)-$\log n$ relation, and its $1\sigma$ dispersion. The magenta vertical lines represent the critical densities of [OIII] (dotted) and [CII] (dashed) transitions. In both panels the two sources with spatial displacement between [CII], [OIII], and UV -- for which our model assumption of a single gas slab could introduce a bias in the derivation of the trends -- are indicated with a different symbol (square).}
    \label{fig:ks_vs_oiiicii}
\end{figure*}
 
 \begin{figure}
     \centering
     \includegraphics[scale=0.5]{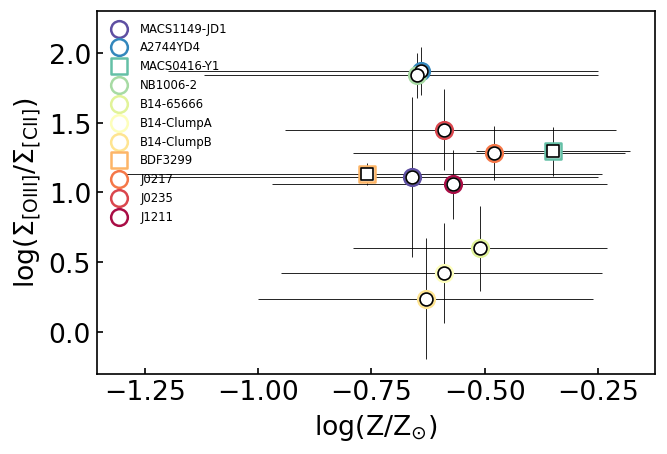}
     \caption{The log(\Soiii/\Scii)-$\log Z$ scatter plot for the high-z sources analyzed in this work. The two sources with spatial di\-spla\-cement between [CII], [OIII], and UV are indicated with a different symbol (square).}
     \label{fig:z_oiiicii}
 \end{figure}

 We conclude that high-$z$ galaxies observed so far seem to show rather extreme ISM properties, namely high-\S*, high \Soiii/\Scii~values. This is expected as they have been selected to be the brightest sources at those redshifts. Finally, if we compare the beam-averaged values of SERRA with the observed ones, we find a very good agreement with the majority of the sources, even though MACS1149-JD1 ($z=9.1$), and AZ744-YD4 ($z=8.38$) have lower \Scii~values than the global values inferred from SERRA at $z=7.7$, and also higher \S*~ when compared with the patches falling in the same range of \Soiii, \Scii. 

\section{Results}
\label{sec:results}
The \Scii, \Soiii and \S*~(along with their uncertainties) in the EoR sample allow us to run our model and derive the posterior probability distributions of the free parameters. The results, along with their $1\sigma$ errors are gathered in Table \ref{tab:higzsources}, while in Appendix \ref{appendixB} we report all the corner plots showing the 2D posterior probability distributions of \nkZ~for each of the galaxy/sub-component analyzed in this work.
\footnote{The current work focuses on a simple situation where the slab geometry, a linear gas-to-dust ratio as a function of metallicity, and a uniform C/O abundance ratio are assumed. The derived parameters can thus have intrinsically larger uncertainties.}
As a first step we focus on to the determination of the burstiness parameter $\kappa_s$.
All the joint [OIII]-[CII] emitters in the EoR sample have $1.0<\log \kappa_s<2.0$, meaning that all of them are starburst galaxies with upwards deviations from the KS relation. These values are also higher than $\log \ks =0.3$ found by \citetalias{vallini2020} for COS3018, a LBG at $z=6.6$, with the same method applied to [CII] and CIII] emission data. While in principle using CIII] or [OIII] should return consistent  \nkZ~values if applied on a galaxy detected in all the three lines, modulo variations with $Z$ of the C/O abundance ratio, dust extinction effects on the CIII] luminosity could decrease $\kappa_s$ if the CIII] value is underestimated.

The $\kappa_s$ derived with our method can be used to constrain the (resolved) gas depletion time $t_{\rm dep}\approx \Sigma_{\rm gas}/\Sigma_{\rm SFR} \propto \kappa_s^{-(1/1.4)} \Sigma_{\rm SFR}^{(1/1.4) -1}$ of the galaxies. High $\kappa_s$ values result in short depletion times, $t_{\rm dep}=6-49\,\rm Myr$. This is in line with resolved $t_{\rm dep}$ values derived in a handful of dusty star forming galaxies (DSFGs) at $z\approx3.5-5$: $t_{\rm dep}=50-200$ Myr \citep{hodge2015}, $t_{\rm dep}=1-30$ Myr \citep{rybak2020}, and $t_{\rm dep}=12-357$ Myr \citep{rizzo2021}. Our analysis thus suggests that joint [CII]-[OIII] emitters, despite not being as dusty as DSFGs, share comparable levels of star formation activity. They also seem to be characterized by ISM conditions favouring an efficient conversion of gas into stars, carving \HII~regions which shine conspicuously in [OIII] emission.

\begin{figure*}
    \centering
    \includegraphics[scale=0.5]{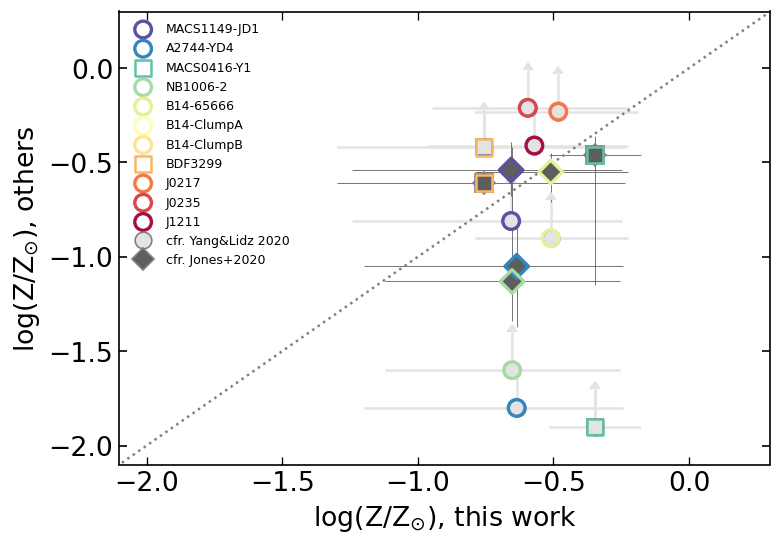}
    \includegraphics[scale=0.5]{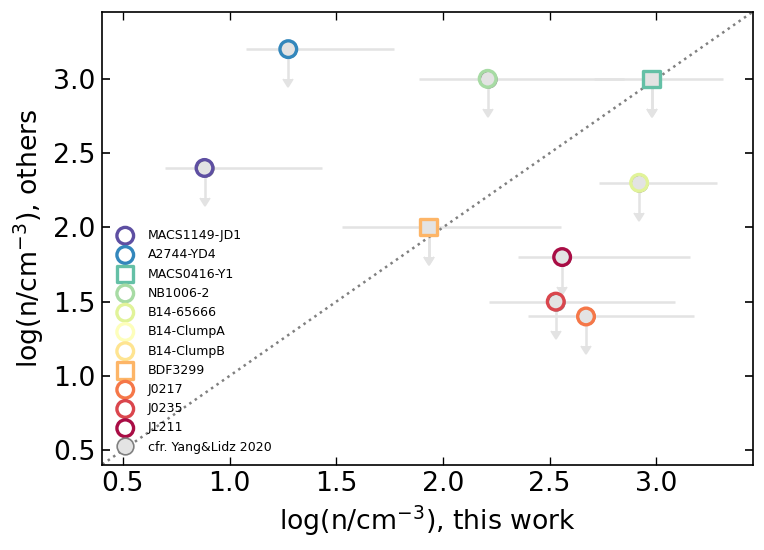}
    \caption{Comparison between metallicity (left panel) and gas density (right panel) derived in this work, and those obtained by \citet{jones2020} (black diamonds) and \citet{yang2020} (grey circles). The dotted line in the two panels represents the 1:1 relation. The two sources with spatial displacement between [CII], [OIII], and UV are indicated with a different symbol (square).}
    \label{fig:cfr_with_others}
\end{figure*}
\subsection{Relative role of \texorpdfstring{($\kappa_s$, Z, n)}{kzn} on  \texorpdfstring{$\Sigma_{\rm [OIII]}/\Sigma_{\rm [CII]}$}{SigmaOIII/SigmaCII}}
To broadly quantify possible trends between \nkZ~and the \Soiii/\Scii, in Fig \ref{fig:ks_vs_oiiicii} we present the best fit linear regression between log(\Soiii/\Scii) and $\log \kappa_s$, $\log n$, $\log Z$, separately. For the burstiness parameter we find the following relation:
\begin{equation}\label{eq:linear_reg_k}
    \log (\Sigma_{\rm [OIII]}/\Sigma_{\rm [CII]}) = \alpha \log \kappa_s + \beta
\end{equation}
with $\alpha= 1.04\pm0.41$, and $\beta =-0.42\pm 0.62$. The correlation is thus linear and can be explained by considering that higher starbursting conditions produce more extended ionized layers \citep{ferrara2019} in which \OIIIeighty~is excited. Eq. \ref{eq:linear_reg_k} can be readily used to estimate the global deviation from the KS relation of EoR galaxies once resolved ALMA [OIII] and [CII] observations are obtained.
%A further interesting possibility is to extend the exploitation of such relation within $\approx 100$ patches of the ISM early galaxies thaks to the help of gravitational lensing.

For what concerns the gas density we obtain:
\begin{equation}\label{eq:linear_reg_n}
  \log (\Sigma_{\rm [OIII]}/\Sigma_{\rm [CII]})  = \alpha \log n  + \beta
\end{equation}
with $\alpha= -0.38\pm 0.21$ and $\beta =2.00 \pm 0.52$. The negative slope can be explained in terms of the different critical density of \OIIIeighty~($n_{\rm c,OIII}=500\,\rm cm^{-3}$ for collisions with free electrons) and [CII] ($n_{\rm c,CII}=3300\,\rm cm^{-3}$, for collisions with neutrals). For galaxies with $n_{\rm c,OIII}< n < n_{c,\rm CII}$ the [OIII]/[CII] luminosity ratio drops, and so does the surface brightness one (see Figure \ref{fig:ks_vs_oiiicii}, right panel). 
%
%The \Soiii/\Scii~ratios of sources with $n<n_{\rm c,OIII}$ are instead less barely influenced by the gas density.
%
If the average density of a galaxy is below $n_{\rm c,OIII}$, and thus also below the $n_{\rm c,CII}$, the gas density does not affect the [OIII]/[CII] ratio anymore leaving the burstiness of the source as the most important physical me\-cha\-nism boosting the [OIII]/[CII] ratio.
Finally,  despite the increasing [OIII]/[CII] trend towards lower $Z$ observed in dwarf galaxies \citep{cormier2015}, we do not find any significant correlation between log(\Soiii/\Scii) and gas metallicity. 

To quantitatively assess the importance of the parameters we also performed a principal component analysis (PCA). PCA is a linear transformation technique by which a set of variables (in our case $\kappa_s$, $n$, $Z$, \Soiii/\Scii) are combined into orthogonal \emph{principal components} (PC). Successive components account for decreasing variance in the sample, under the constraint that each PC is orthogonal to the previous ones.
In order to perform the PCA we first normalize the variables to zero mean in logarithmic space:
\begin{eqnarray}
     x_1 &=& \log(\kappa_s)-\langle\log(\kappa_s)\rangle\\
     x_2 &=& \log(n)-\langle\log(n)\rangle\\
     x_3 &=& \log(Z)-\langle\log(Z)\rangle\\
     x_4 &=& \log(\Sigma_{\rm [OIII]}/\Sigma_{\rm [CII]})-\langle\log(\Sigma_{\rm [OIII]}/\Sigma_{\rm [CII]})\rangle
\end{eqnarray}
and then we derive enough PCs to explain 99\% of the sample variance. In our case we need three components (PC1, PC2, and PC3) which account for 69\%, 28\%, and 2\% of the variance, respectively:
\begin{eqnarray}
     {\rm PC1} &=& 0.04 x_1 - 0.86 x_2 - 0.06 x_3 + 0.50 x_4\\
     {\rm PC2} &=& - 0.60 x_1 - 0.41 x_2 - 0.10 x_3 - 0.68 x_4\\
     {\rm PC3} &=& - 0.80 x_1 + 0.27 x_2 - 0.04 x_3 - 0.54 x_4.
\end{eqnarray}
The first eigenvector, PC1, is dominated by the gas density while PC2 and PC3 are dominated by $\kappa_s$. The metallicity is the least influential parameter in all the PCs. Given that PC3 accounts for only $\sim 2\%$ of the variance it can be exploited for establishing an optimized view of the parameter space defined by $\log \kappa_s$, $\log n$, $\log Z$ and $\log (\Sigma_{\rm [OIII]}/\Sigma_{\rm [CII]}$) thus giving:
\begin{equation}
    \log \left(\frac{\Sigma_{\rm [OIII]}}{\Sigma_{\rm [CII]}}\right)= 0.16 + 1.5 \log \kappa_s - 0.5 \log n + 0.07 \log Z.
\end{equation}
This relation is especially useful from a theoretical stand point, as it combines in a single expression the dependence of the [OIII]/[CII] ratio on the three main physical parameters of the problem. As a caveat we note that, although in general PCA is very effective in isolating correlations among parameters, both the sample size and the metallicity range spanned here are small. Hence, further analysis on larger samples would be beneficial. Alternatively, eq.s \ref{eq:linear_reg_k} and \ref{eq:linear_reg_n}, can be adopted to infer $\kappa_s$ or $n$ from measured \Soiii/\Scii~values in galaxies lacking information on all parameters.

To summarize, our analysis suggests that high \Soiii/\Scii~values so far measured in high-$z$ galaxies are likely produced by ongoing starbursts. Note that also several SED fitting analysis \citep[e.g.][]{laporte2017, roberts-borsani2020, jones2020} and cosmological simulations \citep{pallottini2017b, pallottini:2019, arata2020} point towards EoR sources being characterized by intermittent star formation histories with separated bursts of star formation.
This highlights the importance of using multi-parameter MCMC approaches to derive the metallicity from [OIII]/[CII] ratios (see Figure \ref{fig:z_oiiicii}) and represents an attempt in the direction suggested by \citet{harikane2020}, who pointed out that the metallicity dependence of the [OIII]/[CII] ratio cannot be easily disentangled from variations\footnote{We caution the reader that in this work we have assumed a fixed  O/C abundance. Recent simulations from \citet{arata2020}, suggest that the increasing trend can be ascribed to enhanced O/C ratios at low-$Z$ \citep[e.g.][]{steidel2016, berg2019}.} in the ionization parameter and $n$.

\subsection{Metallicity and density in the EoR sample}
In this section we discuss the major implications of the overall conditions of the ISM in EoR galaxies as inferred from the \Soiii/\Scii~ratios and we compare them with findings of other works in the literature.
 Despite the high redshift range spanned by the sample ($z\approx 6-9$), we find relatively high metallicities, $Z\approx 0.2-0.5\, \rm Z_{\odot}$, and a wide range of densities, $\log(n/{\rm cm^{-3}}) \approx 0.7-3.0$. This seems to indicate that joint [OIII]-[CII] emitters so far detected in the EoR have a fairly enriched ISM with some of the sources (MACS0416-Y1, B14-65666) presenting also high $\log n >2.5$ average densities. These values are in line with results obtained from zoom-in cosmological simulations of galaxies with comparable stellar mass \citep{pallottini2017b, pallottini:2019}.

In Figure \ref{fig:cfr_with_others} we compare our findings with results from two recently published independent analysis \citep{jones2020, yang2020}. 
\citet{jones2020} -- with the aim of studying the mass-metallicity relation \citep[MZR,][and references therein]{maiolino2019} -- derived the (O/H) abundance in all but three sources of our sample. The oxygen abundance was inferred from the \OIIIeighty$/\rm H\beta$ ratio, where the $\rm H\beta$ has been computed from the photometrically-derived SFR of each galaxy. \citet{yang2020}, instead, presented a physically motivated analytical model for the \OIIIeighty~and \OIIIfifty~line emission, relating the \OIIIeighty/SFR ratio of a galaxy to the average electron density and metallicity in the \HII~regions. Due to the $n_e$-$Z$ degeneracy of \OIIIeighty~emissivity equations from \citet{yang2020} they compute the resulting 2D posteriors, in the $n-Z$ plane. They quote the $1\sigma$ lower bound on $Z$, and the $1\sigma$ upper bound on $\log n$ as the most robust results from their analysis, which we therefore consider in our comparison.

The metallicities\footnote{We convert 12+(O/H) to $Z$ assuming $\rm 12+(O/H)_{\odot}=8.69$ \citep{asplund2009}.} derived with the MCMC are in nice agreement with those inferred from \citet{jones2020} which, in line with our previous discussion, pointed out the relatively high values of metallicities and the lack of strong evolution of the MZR towards $z=7-8$. Our $Z$ values also agree within the errors with the lower limits on metallicity from \citet{yang2020}. On the other hand, we exceed their upper bounds on the \HII region electron density ($n_e$) in four out of nine sources

This is expected, as with our model we constrain the mean gas density of two connected environments, namely the \HII regions (dominating the [OIII] emission) and the PDRs (dominating the [CII] emission). The PDRs are generally characterized by higher densities ($\log n = 2 - 4$) with respect to that of \HII regions ($\log n_e =0-3$), as also outlined in the model validation section when comparing $n_{\rm PDR}$ and $n_{\rm HII}$ derived by \citet{cormier2019} in local dwarf galaxies with those inferred with our model. Finally, note that the values of $n$ are in line with the mean gas density in the central regions of the SERRA galaxies characterized by $\log n = 2.5-3.0$ \citep{pallottini:2019}.

\begingroup
\renewcommand{\arraystretch}{1.5}
\begin{table*}
\caption{List of the EoR galaxies analyzed in this work along with the input \Scii, \Soiii, \S*~values and the $Z$, $\kappa_s$ $n$ from the MCMC. We have assumed a fiducial 0.3 dex uncertainty for the \S*. References as follows: M15 \citet{maiolino2015}, I16 \citet{inoue2016}, L17 \citet{laporte2017}, C17 \citet{carniani2017}, H18 \citet{hashimoto2018}, H19 \citet{hashimoto2019}, T19 \citet{tamura2019}, B20 \citet{bakx2020}, HK20 \citet{harikane2020}, C20 \citet{carniani2020}.}              % title of Table
\label{tab:higzsources}   
\centering                                     
\begin{tabular}{lcccccccc}      
\hline\hline                        % inserts double horizontal lines
Name & $z$ & $\log(\frac{\Sigma_{\rm [CII]}}{L_{\odot}\,\rm kpc^{-2}})$ & $\log(\frac{\Sigma_{\rm [OIII]}}{L_{\odot}\,\rm kpc^{-2}})$ & $\log(\frac{\Sigma_{\rm SFR}}{\rm M_{\odot}\rm\, yr^{-1}\, kpc^{-2}})$ & $\log(Z/Z_{\odot})$ & $\log(\kappa_s)$ & $\log(\frac{n}{\rm cm^{-3}})$ & Refs \\
\hline                                   % inserts single horizontal line

MACS1149-JD1 &      9.11 & 5.44$\pm 0.25$ & 6.55$\pm 0.81$ & 0.68 & $-0.66^{+0.41}_{-0.59}$ & $0.96^{+0.13}_{-0.34}$ & 0.88$^{+0.55}_{-0.19}$ & H18, C20\\
A2744-YD4 &         8.38 & 5.62$\pm 0.33$ & 7.49$\pm 0.24$ & 1.40 & $-0.64^{+0.39}_{-0.56}$ & $1.89^{+0.26}_{-0.11}$ & 1.28$^{+0.50}_{-0.20}$ & L17, C20\\
MACS0416-Y1 &       8.31 & 7.27$\pm 1.21$ & 8.57$\pm 0.25$ & 0.67 & $-0.35^{+0.17}_{-0.17}$ & $1.63^{+0.12}_{-0.12}$ & 2.98$^{+0.33}_{-0.27}$ & T19, B20\\
SXDF-NB1006-2 &     7.21 & 6.31$\pm 0.24$ & 8.15$\pm 0.22$ & 0.84 & $-0.65^{+0.40}_{-0.47}$ & 1.65$^{+0.10}_{-0.06}$ & 2.21$^{+0.64}_{-0.32}$ & I16, C20\\
B14-65666       &   7.15 & 7.58$\pm 0.33$ & 8.18$\pm 0.43$ & 1.04 & $-0.51^{+0.28}_{-0.28}$ & 1.31$^{+0.18}_{-0.18}$ & 2.92$^{+0.36}_{-0.19}$ & H19\\
~$\cdots$~ClumpA &   "   & 7.55$\pm 0.38$ & 7.97$\pm 0.50$ & 1.18 & $-0.59^{+0.35}_{-0.36}$ & 1.23$^{+0.21}_{-0.23}$ & 2.88$^{+0.37}_{-0.20}$ & H19\\
~$\cdots$~ClumpB &   "   & 7.70$\pm 0.64$ & 7.94$\pm 0.61$ & 1.46 & $-0.63^{+0.37}_{-0.37}$ & 1.26$^{+0.26}_{-0.24}$ & 2.98$^{+0.26}_{-0.25}$ & H19 \\
BDF3299 &           7.11 & 6.27$\pm 0.13$ & 7.41$\pm 0.11$ & 0.70 & $-0.76^{+0.52}_{-0.54}$ & 1.12$^{+0.08}_{-0.05}$ & 1.93$^{+0.62}_{-0.41}$ & M15, C17 \\
J0217 &             6.20 & 7.19$\pm 0.21$ & 8.47$\pm 0.27$ & 1.28 & $-0.48^{+0.29}_{-0.31}$ & 1.82$^{+0.12}_{-0.11}$ & 2.67$^{+0.50}_{-0.50}$ & HK20\\
J0235 &             6.09 & 6.95$\pm 0.45$ & 8.40$\pm 0.41$ & 1.31 & $-0.59^{+0.38}_{-0.35}$ & 1.88$^{+0.13}_{-0.14}$ & 2.53$^{+0.56}_{-0.31}$ & HK20\\
J1211 &             6.02 & 7.17$\pm 0.23$ & 8.23$\pm 0.35$ & 1.25 & $-0.57^{+0.34}_{-0.40}$ & 1.56$^{+0.17}_{-0.17}$ & 2.56$^{+0.60}_{-0.21}$ & HK20\\
\hline                                             %inserts single line
\end{tabular}
\end{table*}
\endgroup

\section{Conclusions}
\label{sec:conclusions}
In this work we have studied the origin of the high \Soiii/\Scii~ratios observed in ALMA-detected EoR galaxies.
We have used a model that relates the \Soiii/\Scii~and \S*~of the galaxy to three quantities: the deviation from the KS relation, the gas metallicity, and the gas density of early galaxies. We have tested our method on a sample of low-$z$ dwarf sources with measured \Scii, \Soiii, \S*, finding an overall good agreement between our results and previous data in the literature. The method has then been applied to the (only) nine galaxies at $z>6$ jointly detected in [CII] and [OIII]. Our main findings are:
\begin{itemize}
    \item High (\Soiii/\Scii$\approx 30$) ratios are not due to observational biases, but they genuinely arise from the extreme gas conditions prevailing in these early galaxies.
    
    \item According to our model, the observed \Soiii/\Scii~ratios correspond to $\kappa_s = 10-100$, i.e. upward \S*~deviations by 10-100$\times$ from the Kennicutt-Schmidt relation.
    
    \item We use $\kappa_s$ to constrain the (resolved) gas depletion time $t_{\rm dep}=6-49\,\rm Myr$. Such low values are in line with resolved $t_{\rm dep}$ measurements of dusty star forming galaxies at $z\approx3-5$.
    
    \item A principal component analysis suggests that \Soiii/\Scii~mostly depend on $\kappa_s$ with a secondary anticorrelation with gas density.
    
    \item The high $\kappa_s$ (low $t_{\rm dep}$) suggests that [CII]-[OIII] emitters are characterized by ISM conditions favouring an efficient conversion of gas into stars, with starburst episodes producing bright [OIII] emission from \HII~regions and hotter dust temperature as recently found by \citet{sommovigo2021}.
    
    \item Linear regressions between \Soiii/\Scii~and one para\-meter at the time return the following relations: $\Sigma_{\rm [OIII]}/\Sigma_{\rm [CII]} \propto \kappa_s$, and  $\Sigma_{\rm [OIII]}/\Sigma_{\rm [CII]} \propto n^{-0.4}$. However, we do not find evidence for a relation with gas metallicity.
    \item Gas metallicity and density are relatively high ($Z=0.1-0.5 Z_\odot$, and $n =10^{1-3} \rm{cm^{-3}}$) in the nine sample galaxies, also in agreement with findings from \citet{jones2020, yang2020}.
\end{itemize}
 Our method will allow to constrain the star-formation law at sub-kpc scales in early galaxies via \Soiii/\Scii~ratios by boosting ALMA spatial resolution with the help of gravitational lensing.
 
\section*{Acknowledgements}
LV, AF, SC, AP, acknowledge support from the ERC Advanced Grant INTERSTELLAR H2020/740120 (PI: Ferrara). 
Any dissemination of results must indicate that it reflects only the author’s view and that the Commission is not responsible for any use that may be made of the information it contains. 
Partial support from the Carl Friedrich von Siemens-Forschungspreis der Alexander von Humboldt- Stiftung Research Award is kindly acknowledged (AF).
We acknowledge the CINECA award under the ISCRA initiative, for the availability of high performance computing resources and support from the Class B project SERRA HP10BPUZ8F. We acknowledge use of Astropy \citep{astropy2018},  Matplotlib \citep{matplotlib2007}, NumPy \citep{NumPy2020}, Seaborn \citep{seaborn2020}, SciPy \citep{scipy2020}, and Pandas \citep{pandas2020}.
LV thank Laura Sommovigo for fruitful discussions. We thank the reviewer for useful suggestions that improved the quality of the paper.

\section*{Data availability}
We release the code, called GLAM (Galaxy Line A\-naly\-zer with MCMC) for the derivation of the \nkZ~from \Soiii/\Scii~observations on Github at:   \url{https://lvallini.github.io/MCMC_galaxyline_analyzer/}.
The SERRA data used in this article were accessed from the computational resources available to the Cosmology Group at Scuola Normale Superiore, Italy.
\bibliographystyle{mnras}
\bibliography{master_biblio} 
\appendix 
\section{Low-z dwarf galaxy sample: data and parameters}\label{appendixA}
\begin{figure}
    \centering
    \includegraphics[scale=0.5]{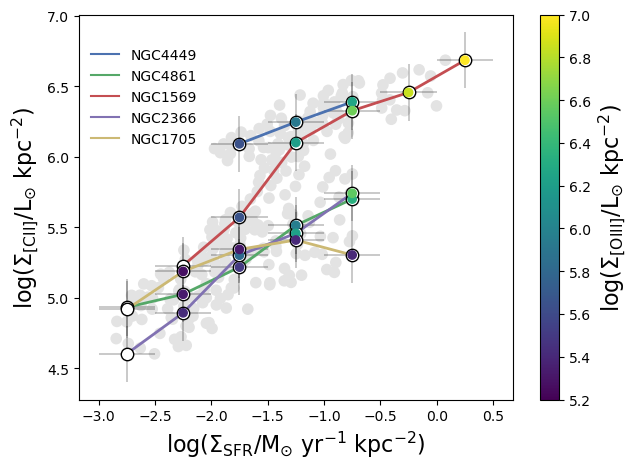}
    \caption{\Scii~vs \S*~relation in spatially resolved dwarf galaxies at $z=0$: NGC 4449 (blue solid line), NGC 4861 (green), NGC 1509 (red), NGC 2366 (purple) and NGC 1705 (yellow). Data (in grey) are taken from \citet{delooze2014} and aggregated in 0.5 dex wide \S*~bins over which we compute the mean \Scii. Points are color-coded as a function of the mean \Soiii~value over the \S*~bins. Those \S*~bins for which the \Soiii~is undetected are shown in white.}
    \label{fig:dgs_resolved}
\end{figure}

 Figure \ref{fig:dgs_resolved} shows the \S*-\Scii~relation for the five dwarf galaxies considered in this work (grey points). For the five galaxies we highlight the \S*~bins over which we compute the average \Soiii~and \Scii~(colored big points with error bars) used in our method validation at low-$z$. Furthermore, in Tab. \ref{tab:metals}, we list the metallicity values of dwarf galaxies from literature against which we test our validation. As a caveat note that different methods were used for deriving the metallicities referenced in Table \ref{tab:metals}. More precisely, \citet{james2016} derived the gas-phase metallicity with narrow-band images based on the R23 method which is offset from the direct-temperature method by 0.23 dex. \citet{annibali2015} used the multi-object slit spectroscopy and obtain the gas-phase metallicity with the direct-temperature method. \citep{madden2013} derived the metallicity using the R23 method, while \citet{mccormick2018} have obtained metallicity using the \citet{pettini2004} method based on the ([OIII]/H$\beta$)/([NII]/H$\alpha$) ratio.
 
 \begin{table}
    \centering
    \begin{tabular}{l|c|l|}
    \hline
         Galaxy &  12+(O/H) & Reference\\
    \hline
         NGC 4449 & $8.20 \pm 0.11$ & \citet{madden2013}\\
         NGC 4861 & $7.89 \pm 0.01$ & \citet{madden2013}\\
         NGC 1569 & $8.16 \pm 0.10$ & \citet{mccormick2018}\\
         NGC 2366 & $8.04 \pm 0.11$ & \citet{james2016}\\
         NGC 1705 & $7.91 \pm 0.08$ & \citet{annibali2015}\\
         \hline
         \end{tabular}
    \caption{Oxygen abundance of the five dwarf galaxies considered in this work.}
    \label{tab:metals}
\end{table}
 
\section{The posterior probability distributions}\label{appendixB}
\begin{figure*}
    \centering
    \includegraphics[scale=0.6]{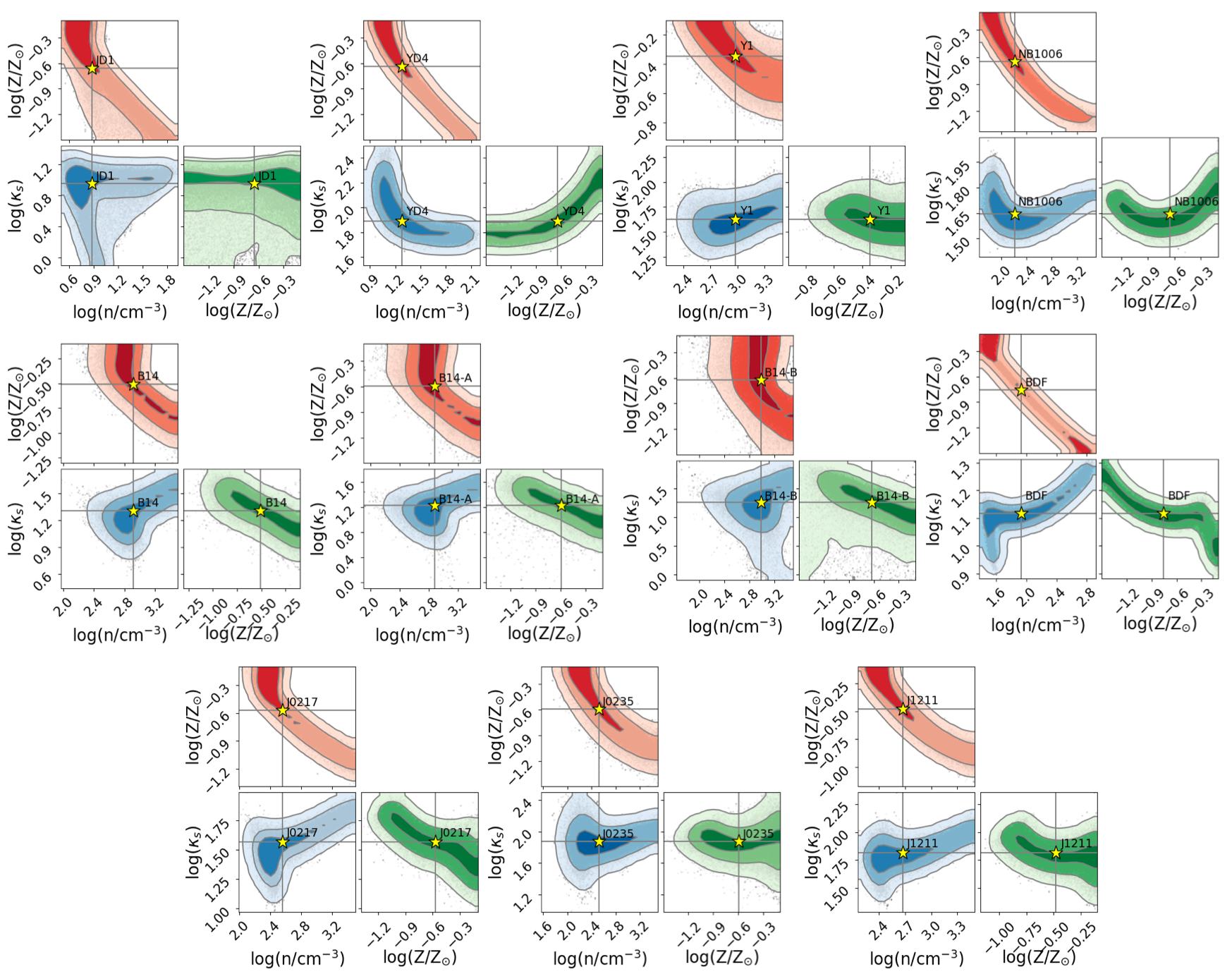}
    \caption{Corner plots with the results of the MCMC for all the galaxies analyzed in this work. The name of each galaxy is indicated with a label close to the best fit location (yellow star) in each plot.}
    \label{fig:mcmc_corners}
\end{figure*}

In Figure \ref{fig:mcmc_corners} we show for each galaxy/sub-component the corner plot of the 2D likelihood distributions for three parameters \nkZ~entering in our analytical models. Contours represent the 1$\sigma$, 2$\sigma$, 3$\sigma$ confidence levels.
%%%%%%%%%%%%%%%%%%%%%%%%%%%%%%%%%%%%%%%%%%%%%%%%%%
% Don't change these lines
%\bsp	% typesetting comment
%\label{lastpage}
\end{document}